\documentclass[aps,pre,superscriptaddress,twocolumn,showpacs]{revtex4}
\usepackage{epsfig}
\usepackage{graphicx}% Include figure files
\usepackage{dcolumn}% Align table columns on decimal point
\usepackage{bm}% bold math
\usepackage{amsmath,amssymb,times}

\begin{document}

\title{Traffic-driven epidemic spreading in correlated networks}

\author{Han-Xin Yang}\email{hxyang01@gmail.com}
\affiliation{Department of Physics, Fuzhou University, Fuzhou
350108, China}

\author{Ming Tang}\email{tangminghuang521@hotmail.com}
\affiliation{Web Sciences Center, University of Electronic Science
and Technology of China, Chengdu 610051, China}

\author{Ying-Cheng Lai}
\affiliation{School of Electrical, Computer and Energy
Engineering, Arizona State University, AZ 85287, USA}

\begin{abstract}

In spite of the extensive previous efforts on traffic dynamics and
epidemic spreading in complex networks, the problem of
traffic-driven epidemic spreading on {\em correlated} networks has
not been addressed. Interestingly, we find that the epidemic
threshold, a fundamental quantity underlying the spreading dynamics,
exhibits a non-monotonic behavior in that it can be minimized for
some critical value of the assortativity coefficient, a parameter
characterizing the network correlation. To understand this
phenomenon, we use the degree-based mean-field theory to calculate
the traffic-driven epidemic threshold for correlated networks. The
theory predicts that the threshold is inversely proportional to the
packet-generation rate and the largest eigenvalue of the betweenness
matrix. We obtain consistency between theory and numerics. Our
results may provide insights into the important problem of
controlling/harnessing real-world epidemic spreading dynamics driven
by traffic flows.
\end{abstract}

\date{\today}


\maketitle

\section{Introduction}

Both epidemic spreading~\cite{1,2,3,4,5,6,7,8,9,10,11} and traffic
transportation~\cite{12,13,14,15,16,17,18} are two types of
fundamental dynamical processes on complex networks. The past
fifteen years have witnessed a great deal of effort in exploring and
understanding these processes. Intuitively, spreading should be
greatly affected by the transportation dynamics as the traffic flows
determine the paths along which a virus or information can
propagate. Nevertheless, for a long time, the two types of dynamical
processes were studied separately. The first series of works to
incorporate traffic dynamics into epidemic spreading were done using
the metapopulation model~\cite{m1,m2,m3,m4,m4.1,m5.0,m5}. In
particular, consider a spatially extended system of a large number
of individuals. In a metapopulation model, all the individuals are
grouped into a number of spatially structured subpopulations
interacting with each other in a manner that can be described as a
network. The links in the network represent the traveling paths of
individuals across different subpopulations. An infected individual
can infect other individuals in the same subpopulation. The
metapopulation model has been used to simulate the spread of human
and animal diseases (such as SARS and H1N1) among different cities.
In 2009, Meloni et al. proposed another traffic-driven epidemic
spreading model~\cite{Meloni}, where each node corresponds to a
router (e.g., in a computer network) and the virus can spread among
nodes through the transmission of packets. A susceptible node will
be infected with certain probability every time it receives a packet
from an infected neighboring node. This model is quite suitable for
studying computer virus propagation.

The model of Meloni et al.~\cite{Meloni} has become a prototypical
framework to address a variety of issues in traffic-driven epidemic
spreading. For traffic flow dynamics on complex networks, an
essential ingredient is the routing strategy. In the model, one can
demonstrate that epidemic spreading can be modulated or controlled
through a proper choice of the local routing strategy~\cite{yang1}
or global routing protocol~\cite{yang3}. For a fixed routing
strategy, the network structure can also affect the spreading
dynamics. For example, by increasing the average network
connectivity, one can effectively delay epidemic
outbreaks~\cite{yang4}. Furthermore, the epidemic threshold can be
enhanced by deliberately removing a subset of links associated with
the large-degree nodes or the edges with the largest algorithmic
betweenness~\cite{yang5}.

In previous studies of traffic-driven epidemic spreading, the
networks were assumed to be uncorrelated. In such a network, the
average degree of the neighboring nodes is uniform across nodes.
However, many real-world networks display various degrees of mixing
patterns~\cite{newman}. Qualitatively, a network is assortatively
(disassortatively) mixing if high-degree nodes tend to connect with
high-degree (low-degree) nodes. Quantitatively, degree mixing can be
characterized by the assortativity coefficient~\cite{newman}:
\begin{equation} \label{eq1}
c=\frac{M^{-1}\sum_{i}j_{i}k_{i}-[M^{-1}\sum_{i}\frac{1}{2}
(j_{i}+k_{i})]^{2}}{M^{-1}\sum_{i}\frac{1}{2}(j_{i}^{2}+k_{i}^{2})
-[M^{-1}\sum_{i}\frac{1}{2}(j_{i}+k_{i})]^{2}},
\end{equation}
where $j_{i}$, $k_{i}$ are the degrees of the nodes at the ends of
the $i$th edge, $M$ is the number of edges in the network, and $i =
1,\ldots,M$. For standard network models such as the
Erd\"{o}s-R\'{e}nyi random graphs~\cite{er} and the
Barab\'{a}si-Albert scale-free networks~\cite{BA}, the assortativity
coefficient $c$ is zero, indicating complete lack of degree
correlation. For many social networks, the assortativity coefficient
$c$ is positive. However, technological and biological networks tend
to be disassortative with negative values of $c$.

In this paper, we study quantitatively how degree mixing, the most
pronounced feature of correlated networks, affects traffic-driven
epidemic spreading. Interestingly, we find a non-monotonic behavior
in that the threshold can be minimized for some critical value of
the assortativity coefficient. To understand this phenomenon, we
employ a degree-based mean-field theory, which allows us to
calculate the traffic-driven epidemic threshold in correlated
networks. We find that the threshold is inversely proportional to
the packet-generation rate - a key parameter characterizing the
traffic dynamics, as well as the largest eigenvalue of the
betweenness matrix. Moreover, the epidemic threshold is negatively
correlated with the maximum algorithmic betweenness. Our theory
predicts the existence of a critical value of the assortativity
coefficient at which the epidemic threshold reaches its minimum,
which agrees with the results from direct numerical simulations.

\section{Correlated scale-free network and traffic-driven epidemic
spreading models} \label{sec:model}

\paragraph*{Construction of scale-free networks with tunable
assortativity coefficient.} We first generate an ensemble of
scale-free networks according to the uncorrelated configuration
model~\cite{UCM}. The steps are: (i) assign to each node $i$, in a
set of $N$ initially disconnected nodes, a number $k_i$ of stubs,
where $k_i$ is drawn from the probability distribution $P(k)\sim
k^{-\gamma}$ and subject to the constraints $k_{min} \leq k_i \leq
\sqrt{N}$ (here we set $k_{min}=2$ and $N=4900$) and $\sum_{i}k_i$
being even, and (ii) generate a network by randomly choosing stubs
and connecting them to form edges, taking into account the
preassigned degrees and avoiding multiple and self-connections. We
then use the algorithm in Ref.~\cite{xs} to obtain networks with
desired degree mixing patterns. Specifically, in order to obtain an
assortative network, at each step we randomly choose two different
edges with four different ends, and then purposefully swap the two
edges by linking the vertices with higher degrees and lower degrees,
respectively. Repeating this procedure while forbidding multiple
connections and disconnected components, we generate a network with
certain degree assortativity without altering the degree
distribution of the original network. Through the opposite operation
that one edge links the highest and the lowest nodes and the other
edge connects the two remaining nodes, we can obtain networks with
disassortative mixing. When the assortativity coefficient of a
network reaches some expected value, we stop the edge-swapping
process.

\paragraph*{Traffic-driven epidemic spreading model.} Following the
pioneering work of Meloni et al.~\cite{Meloni}, we incorporate the
traffic dynamics into the standard susceptible-infected-susceptible
epidemic spreading model~\cite{SIS}, as follows. At each time step,
$R$ new packets are generated with randomly chosen source and
destination nodes, and each node $i$ can deliver at most $D_{i}$
packets toward their destinations. (For simplicity, we set $D_{i}$
to be infinite.) Packets are forwarded according to a given routing
algorithm (we use the shortest-path routing scheme). The queue
length of each node is assumed to be unlimited, to which the
first-in-first-out principle is applied. Each newly generated packet
is placed at the end of the queue of its source node. Once a packet
reaches its destination, it is removed from the system. A node can
be in two discrete states: susceptible or infected. After a
transient time, the total number of delivered packets at each time
step will reach a steady value. Subsequently, an initial fraction of
nodes $\rho_{0}$ is set to be infected (we choose $\rho_{0}=0.1$ in
numerical experiments). Differing from the conventional spreading
dynamics where there is a certain probability that a node can be
infected if one or more of its neighboring nodes are infected, here
the infection spreads in the network {\em through packet exchanges}.
In particular, all packets queuing at an infected node are infected,
while all packets at a susceptible node are uninfected. A
susceptible node has the probability $\beta$ of being infected every
time it receives an infected packet from any infected neighboring
nodes. With probability $1-\beta$, the virus in an infected packet
will be cleaned by some anti-virus software (say, in a computer
network) in the susceptible node. The infected nodes are recovered
at the rate $\mu$ (we set $\mu=1$).

\section{Numerical results and mean-field theory} \label{sec:results}

\begin{figure}
\begin{center}
 \scalebox{0.4}[0.4]{\includegraphics{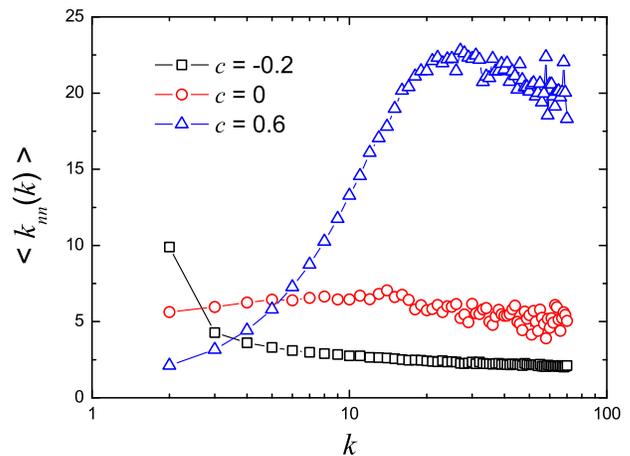}} \caption{(Color online)
For scale-free networks of degree exponent $\gamma=3$ with different
values of $c$, the average degree of the nearest neighbors, $\langle
k_{nn}(k) \rangle$, as a function of the degree $k$. Each data point
is the result of averaging over 100 independent network
realizations. For assortative networks ($c > 0$), $\langle k_{nn}(k)
\rangle$ is an increasing function of $k$ (for small $k$ values).
For disassortative networks ($c < 0$), $\langle k_{nn}(k) \rangle$
is a decreasing function of $k$. For uncorrelated networks ($c =
0$), $\langle k_{nn}(k) \rangle$ is independent of $k$.}
\label{fig:knn}
\end{center}
\end{figure}

\paragraph*{Numerical results of the dependence of the epidemic
threshold on the assortativity coefficient.} A useful quantity to
characterize correlated networks is the average degree of the
nearest neighbors, denoted as $\langle k_{nn}(k) \rangle$.
Figure~\ref{fig:knn} shows $\langle k_{nn}(k) \rangle$ as a function
of degree $k$ for different values of the assortativity coefficient
$c$. From Fig.~\ref{fig:knn}, one can see that $\langle k_{nn}(k)
\rangle$ is almost independent of $k$ when $c=0$. For assortative
(disassortative) networks, $\langle k_{nn}(k) \rangle$ is an
increasing (decreasing) function of $k$, indicating that
large-degree nodes tend to connect with large(small)-degree nodes.

\begin{figure}
\begin{center}
 \scalebox{0.4}[0.4]{\includegraphics{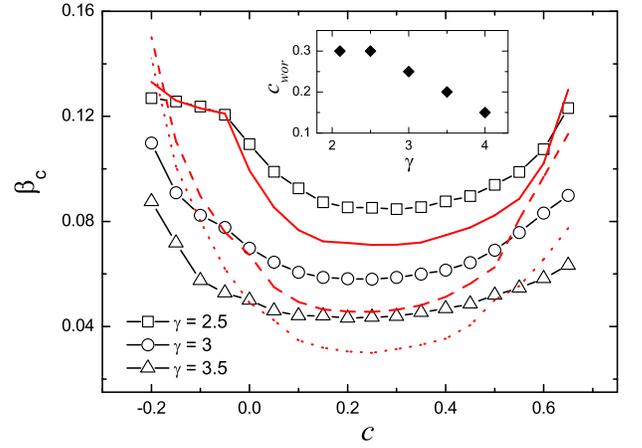}}
\caption{(Color online) For scale-free networks of different values
of degree exponent $\gamma$, epidemic threshold $\beta_{c}$ as a
function of the assortativity coefficient $c$. The solid, dashed,
and dotted curves are theoretical predictions for the cases $\gamma$
= 2.5, 3, and 3.5, respectively. The inset shows the critical value
$c_{wor}$ as a function of $\gamma$. The packet-generation rate is
$R=1500$. Each data point is result of averaging over 100 random
network realizations.} \label{fig:beta_c}
\end{center}
\end{figure}

A fundamental quantity in SIS dynamics is the epidemic threshold
$\beta_{c}$, below which the epidemic dies off
asymptotically~\cite{threshold1,threshold2}. Figure~\ref{fig:beta_c}
shows $\beta_{c}$ versus the assortativity coefficient $c$ for
different values of the scale-free degree exponent $\gamma$. We see
a non-monotonic behavior. Especially, for each value of $\gamma$
simulated, there exists a critical value of $c$, denoted as
$c_{wor}$, at which the epidemic threshold is minimized. The inset
of Fig.~\ref{fig:beta_c} shows that $c_{wor}$ decreases from 0.3 to
0.15 as $\gamma$ increases from 2.1 to 4.

\paragraph*{Mean-field analysis of epidemic threshold for
correlated networks.} We seek to understand the counterintuitive
behavior exemplified in Fig.~\ref{fig:beta_c}. The base of our
analysis is the standard degree-based mean-field
theory~\cite{theory1} for complex networks. First, we write the
dynamical rate equations for the traffic-driven SIS model as
\begin{equation} \label{eq2}
\frac{d\rho_{k}(t)}{dt}=-\rho_{k}(t)+\beta k R [1-\rho_{k}(t)]
\sum_{k^{'}}b_{k^{'}\rightarrow{k}}P(k^{'}|k)\rho_{k^{'}}(t),
\end{equation}
where $\rho_{k}$ is the density of the infected nodes of degree $k$,
$P(k^{'}|k)$ is the conditional probability that a node of degree
$k$ is connected to a node of degree $k^{'}$, and
$b_{k^{'}\rightarrow{k}}$ is the algorithmic
betweenness~\cite{theory2} of a directed edge from a node of degree
$k^{'}$ to a node of degree $k$. The first term in Eq.~(\ref{eq2})
is the recovery rate of the infected nodes, and the second term
takes into account the probability that a node of degree $k$ belongs
to the susceptible class of fraction $1-\rho_{k}(t)$ and gets
infected via packets from some infected nodes. The algorithmic
betweenness of a directed edge, $b_{i\rightarrow j}$, is the number
of packets passing from node $i$ to its neighboring node $j$ each
time if the packet-generation rate is $R=1$. For an undirected
network, we have $b_{i\rightarrow j}=b_{j\rightarrow i}$. For the
shortest-path routing protocol, the algorithmic betweenness of a
directed edge $b_{i\rightarrow j}$ can be calculated through
\begin{equation}\label{eq3}
b_{i\rightarrow j}=\frac{1}{N(N-1)}\sum_{m\neq n}
\frac{\sigma_{m\rightarrow n}(i\rightarrow j)}{\sigma_{m\rightarrow
n}},
\end{equation}
where $\sigma_{m\rightarrow n}$ is the total number of the shortest
paths going from node $m$ to $n$, and $\sigma_{m\rightarrow
n}(i\rightarrow j)$ is the number of shortest paths going from $m$
to $n$ and passing from node $i$ to its neighboring node $j$. The
algorithmic betweenness of a node $i$ is given by
$b_{i}=\sum_{j\in\Omega_{i}}b_{j\rightarrow i}$, where the summation
runs over the nearest neighbor set $\Omega_{i}$ of node $i$.

\begin{figure}
\begin{center}
 \scalebox{0.4}[0.4]{\includegraphics{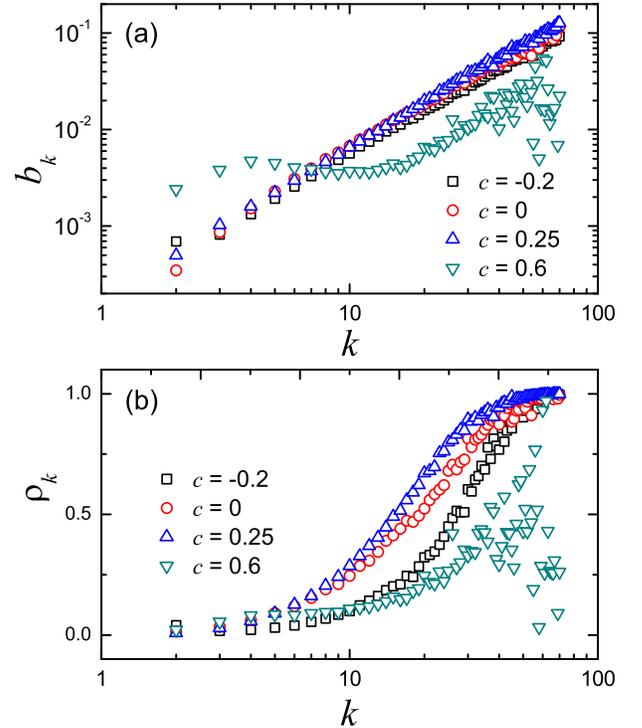}}
 \caption{(Color online) For scale-free networks of degree exponent
$\gamma = 3$ and different values of the assortativity coefficient
$c$, (a) algorithmic betweenness $b_k$ and (b) density of infected
nodes $\rho_k$ as a function of the degree $k$. For all cases, the
packet-generation rate is $R=1500$ and the density of the infected
nodes in the whole population is $\rho\approx 0.04$. Each data point
results from averaging over 100 random network realizations.}
 \label{fig:b_k}
\end{center}
\end{figure}

\begin{figure*}
\begin{center}
 \scalebox{0.7}[0.7]{\includegraphics{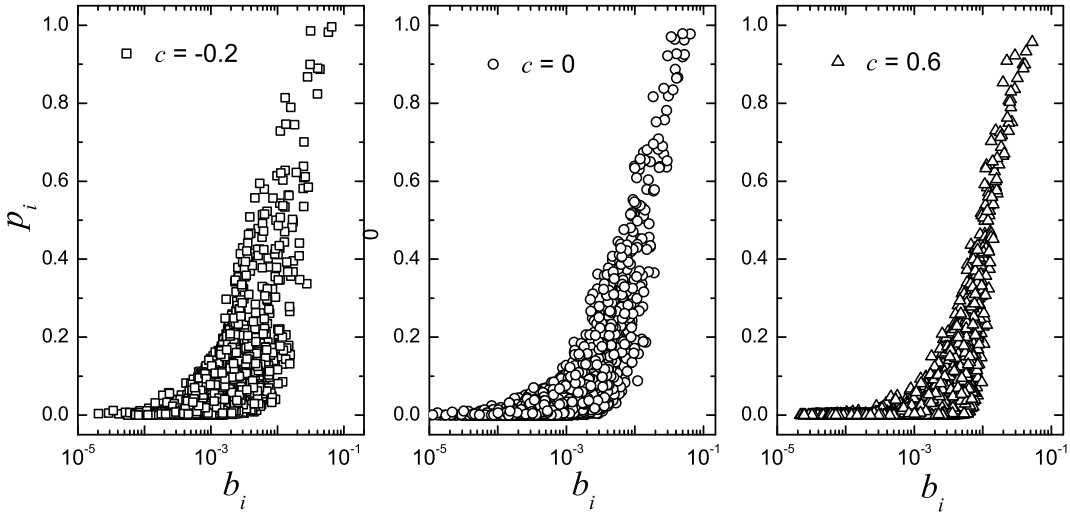}}
 \caption{For scale-free networks of degree exponent
$\gamma = 3$, dependence of the infection probability $p_{i}$ on the
algorithmic betweenness $b_{i}$ for different nodes. The values of
the assortativity coefficients $c$ are -0.2, 0 and 0.6 for the left,
middle and right panels respectively. Other parameters are:
packet-generation rate $R=1500$ and density of infected nodes
$\rho\approx 0.04$.} \label{fig:p_i}
\end{center}
\end{figure*}

For small values of $\rho_{k}$, we can linearize Eq.~(\ref{eq2}) to
obtain
\begin{equation}\label{eq4}
\frac{d\rho_{k}(t)}{dt}\simeq \sum_{k^{'}}L_{kk^{'}}\rho_{k^{'}}(t),
\end{equation}
where the Jacobian matrix element is
\begin{equation}\label{eq5}
L_{kk^{'}}=-\delta_{kk^{'}}+\beta k R
P(k^{'}|k)b_{k^{'}\rightarrow{k}},
\end{equation}
and $\delta_{kk^{'}}$ is the Kronecker delta function. The solution
$\rho_{k} = 0$ will be unstable if there exists at least one
positive eigenvalue of the Jacobian matrix $\textbf{L}$. The endemic
phase ($\rho_{k} >0$)  will thus take place for $-1+\beta R
\Lambda_{m}> 0$, where $\Lambda_{m}$ is the largest eigenvalue of
the {\em betweenness matrix} $\textbf{C}$ with its elements given by
\begin{equation}\label{eq6}
C_{kk^{'}}= k P(k^{'}|k)b_{k^{'}\rightarrow{k}}.
\end{equation}
The element $C_{kk^{'}}$ gives the number of packets that a node of
degree $k$ receives from all neighboring nodes of degree $k^{'}$ at
each time step when the packet-generation rate $R=1$. Since
$\textbf{C}$ is non-negative and irreducible, the Perron-Frobenius
theorem~\cite{theory4} stipulates that its largest eigenvalue is
real and positive. The endemic state then occurs for
\begin{equation}\label{eq7}
\beta> \beta_{c}=\frac{1}{R\Lambda_{m}}.
\end{equation}

For uncorrelated networks, we can use two approximations: ($i$)
$P(k^{'}|k)\propto k^{'}P(k^{'})$, and ($ii$) $b_{k^{'}\rightarrow
k} \propto b_{k}b_{k^{'}}/kk^{'}$, where the latter allows us to
identify the key parameters of the underlying dynamics. Under these
approximations, the element of the betweenness matrix can be reduced
to
\begin{displaymath}
C_{kk^{'}}^{uc} \approx b_{k^{'}}b_{k} P(k^{'})/ \langle b \rangle,
\end{displaymath}
which has a unique non-zero eigenvector with the eigenvalue
$\Lambda_{m}^{uc}=\langle b^{2} \rangle/ \langle b \rangle$. The
epidemic threshold for uncorrelated networks can then be obtained as
\begin{equation}\label{eq8}
\beta_{c}^{uc}=\frac{ \langle b \rangle}{R\langle b^{2} \rangle},
\end{equation}
where $b$ represents the algorithmic betweenness of a node. This
result was obtained previously by a slightly different version of
the mean-field theory (heterogeneous mean-field
theory)~\cite{Meloni}.

To validate our theoretical results [Eq.~(\ref{eq7})], we compare
them with numerically calculated dependence of $\beta_{c}$ on the
assortativity coefficients $c$. As shown in Fig.~\ref{fig:beta_c},
the theoretical and numerical results are consistent.

How degree mixing affects the algorithmic betweenness and the
infection probability for nodes of different degrees? The nodal
algorithmic betweenness within the degree class $k$ is defined as
$b_k=\Sigma_{i}b_{i}/N_{k}$, where $N_{k}$ is the number of nodes
within each degree class $k$ and the summation runs over all nodes
of degree $k$. Figure~\ref{fig:b_k}(a) shows the dependence of $b_k$
on degree $k$. We see that for dissortative (e.g., $c=-0.2$),
uncorrelated ($c=0$), and weakly assortative (e.g., $c=0.25$)
networks, $b_k$ increases with $k$, and the scaling of $b_k$ with
$k$ follows a power law. However, for strongly assortative (e.g.,
$c=0.6$) networks, the scaling is not as straightforward in that the
values of $b_k$ for large-degree nodes are not necessarily high but
instead spread in a relatively wide range. (Similar phenomena were
observed previously in some real-world networks~\cite{bjk}.) The
density of the infected nodes within the degree class $k$ is
$\rho_k=I_{k}/N_{k}$, where $N_{k}$ and $I_{k}$ are the numbers of
nodes and of infected nodes within the degree class, respectively.
From Fig.~\ref{fig:b_k}(b), we see that for dissortative (e.g.,
$c=-0.2$), uncorrelated ($c=0$), and weakly assortative (e.g.,
$c=0.25$) networks, $\rho_k$ increases with $k$. However, for
strongly assortative (e.g., $c=0.6$) networks, the value of $\rho_k$
for large-degree nodes can assume relatively small values.

\begin{figure}
\begin{center}
 \scalebox{0.4}[0.4]{\includegraphics{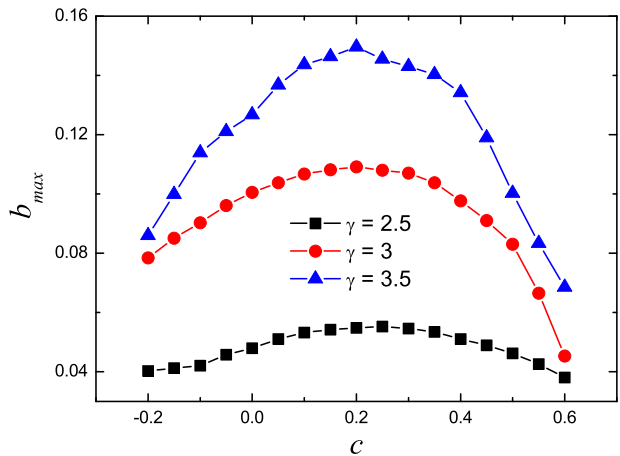}} \caption{(Color online) For scale-free networks of different values of
the degree exponent $\gamma$, the maximum nodal algorithmic
betweenness $b_{max}$ versus the assortativity coefficient $c$. Each
data point is result of averaging over 100 random network
realizations.} \label{fig:b_max}
\end{center}
\end{figure}

\begin{figure}
\begin{center}
 \scalebox{0.4}[0.4]{\includegraphics{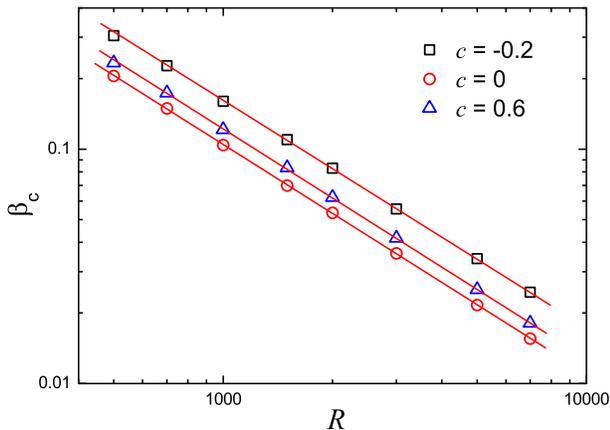}} \caption{(Color online) For scale-free networks of degree exponent
$\gamma = 3$ with different values of the assortativity coefficient
$c$, epidemic threshold $\beta_{c}$ as a function of the
packet-generation rate $R$. The slope of the fitted line is about
-1. Each data point results from an average over 100 network
realizations.}
 \label{fig:beta_c_R}
\end{center}
\end{figure}

Combining Figs.~\ref{fig:b_k}(a) and ~\ref{fig:b_k}(b), we deduce
that the algorithmic betweenness of a node is positively correlated
with the probability of its being infected. We define the infection
probability of a node $i$ as $p_{i}=T_{I}/T$, where $T$ is the total
time lapse and $T_{I}$ is the number of time steps at which node $i$
is infected (here we set $T=1000$). In Fig.~\ref{fig:p_i}, we plot
the dependence of $p_{i}$ on $b_{i}$ for different values of $c$. We
see that, as a general trend, $p_{i}$ increases with $b_{i}$ and the
largest $b_{i}$ corresponds to the highest value of $p_{i}$,
regardless of the value of $c$.

Since the node having the maximum algorithmic betweenness $b_{max}$
is most likely to be infected, it plays a crucial role in
determining the traffic-driven epidemic threshold. In particular,
only when the spreading rate is large enough to successfully infect
the node with $b_{max}$ will the epidemic spreading be able to
sustain. To confirm this, we calculate the dependence of $b_{max}$
on the assortativity coefficient $c$ for networks with different
values of the degree exponent $\gamma$, as shown in
Fig.~\ref{fig:b_max}. We see that, for each value of $\gamma$,
$b_{max}$ peaks at some moderate value of $c$. Comparing with the
results in Fig.~\ref{fig:beta_c}, we see that the $b_{max}$ versus
$c$ behavior is opposite to that of $\beta_c$ versus $c$, implying
that increasing $b_{max}$ causes $\beta_{c}$ to decrease. The
critical value of $c$ leading to the maximum $b_{max}$ is in fact
indistinguishable from the value that minimizes $\beta_{c}$. As a
by-product, the phenomenon shown in Fig.~\ref{fig:b_max} suggests a
relationship between the maximum algorithmic betweenness and the
largest eigenvalue of the betweenness matrix, for which a
mathematical understanding is not yet available.

Finally, we study the epidemic threshold $\beta_{c}$ as a function
of the packet-generation rate $R$ for different values of the
assortativity coefficient $c$. The results are shown in
Fig.~\ref{fig:beta_c_R}. We see that $\beta_{c}$ scales inversely
with $R$ for each value of $c$, as predicted by our mean-field
theory [Eq.~(\ref{eq7})]. This is intuitively correct, as an
increase in the traffic flow generally facilitates epidemic
outbreak.

\section{Conclusions}

To summarize, our investigation of traffic-driven epidemic spreading
in correlated scale-free networks reveals the existence of a
critical level of network correlation for which the outbreak of
epidemic is maximally promoted. We use an extended degree-based
mean-field theory, taking into account the traffic flow dynamics, to
account for the phenomenon. In addition, we find that nodes with
larger algorithmic betweenness are more likely to be infected. We
note that in traditional epidemic spreading where infections are
transmitted as a reactive process from nodes to all neighbors, the
epidemic threshold is largely determined by the nodes with the
largest degree~\cite{Castellano}. Our results reveal, however, that
for traffic-driven epidemic spreading, the threshold mainly depends
on the node with the maximum algorithmic betweenness. In particular,
the threshold tends to decrease with the value of the maximum
algorithmic betweenness. While for uncorrelated or disassortative
networks, the largest-degree node typically has the maximum
algorithmic betweenness, this is not the case for assortative
networks. Our results indicate that in traffic-driven spreading
dynamics, the structural properties of the network can have a
significant effect on epidemic spreading and outbreak, providing a
potential mechanism to control various spreading dynamics in
real-world situations.

\begin{acknowledgments}

This work was supported by the National Science Foundation of China
(Grants No.~61403083 and No.~91324002), and the Natural Science
Foundation of Fujian Province, China (Grant No.~2013J05007). YCL was
supported by the Army Research Office (ARO) under Grant
No.~W911NF-14-1-0504.

\end{acknowledgments}

\end{document}